\documentclass[aps,prd,twocolumn,preprintnumbers,nofootinbib,superscriptaddress,amsmath]{revtex4-2}
    
    \usepackage{graphicx}
    \usepackage[caption=false]{subfig}
    \usepackage[usenames,dvipsnames]{xcolor} 
    \usepackage{pgfplots}
    \usepackage[utf8]{inputenc}
    \usepackage{color}
    \usepackage{hyperref}
    \usepackage[normalem]{ulem} 
    \usepackage{physics}
    \usepackage{enumitem}
    \usepackage{comment}
    \usepackage{bm}
    \usepackage{aas_macros}
    \usepackage[capitalise]{cleveref}
    \usepackage{appendix}
    \usepackage{multirow}
    \usepackage{lipsum}

\hypersetup{
  breaklinks = true,
  colorlinks   = true, 
  urlcolor     = blue, 
  linkcolor    = blue, 
  citecolor   = blue 
}
    \allowdisplaybreaks[1]
    \graphicspath{{figures_revised/}{figures/}}

\usepackage{float}

\newcommand{\splitatcommas}[1]{%
  \begingroup
  \begingroup\lccode`~=`, \lowercase{\endgroup
    \edef~{\mathchar\the\mathcode`, \penalty0 \noexpand\hspace{0pt plus 1em}}%
  }\mathcode`,="8000 #1%
  \endgroup
}

\begin{document}


\title{AdS - de Sitter transition in interacting dark energy models}

\author{Y. Zhao}
\email{yifeizhao@on.br}
\affiliation{Observatório Nacional, Rio de Janeiro - RJ, 20921-400, Brazil}

\author{J. S. Alcaniz}
\email{alcaniz@on.br}
\affiliation{Observatório Nacional, Rio de Janeiro - RJ, 20921-400, Brazil}

\author{H. A. Borges}
\email{humberto@ufba.br}
\affiliation{Instituto de F\'isica, Universidade Federal da Bahia, Salvador - BA, 40210-340, Brazil}

\author{S. Carneiro}
\email{saulocarneiro@on.br}
\affiliation{Observatório Nacional, Rio de Janeiro - RJ, 20921-400, Brazil}
\affiliation{Instituto de F\'isica, Universidade Federal da Bahia, Salvador - BA, 40210-340, Brazil}

\author{R. von Marttens}
\email{rodrigovonmarttens@gmail.com}
\affiliation{Instituto de F\'isica, Universidade Federal da Bahia, Salvador - BA, 40210-340, Brazil}
\affiliation{PPGCosmo, Universidade Federal do Espírito Santo, Vitória-ES, 29075-910, Brasil}

\date{\today}

\begin{abstract}
Recent cosmological results, including the persistent $H_0$ tension and indications of dynamical dark energy, have raised the possibility that the standard $\Lambda$-Cold Dark Matter ($\Lambda$CDM) model may be incomplete. At the same time, theoretical developments based on swampland conjectures indicate that the realization of a stable de Sitter vacuum within string theory is challenging. Motivated by these observational and theoretical considerations, we investigate an interacting $\Lambda$ scenario characterized by a general linear coupling in the dark sector. At the background level, this model naturally allows the dark energy density to evolve from negative to positive values, opening the possibility of a transition from an anti-de Sitter-like phase to the de Sitter-like phase inferred by current observations. We derive the analytic background solutions and perturbation equations, discuss the physically acceptable region of parameter space, and constrain the model with DES-Dovekie type Ia supernovae, DESI BAO and CMB data. Our analysis suggests a parameter space allowing the dark energy density to change sign while maintaining positive dark matter density, motivating further investigation of an interaction-driven anti-de Sitter to de Sitter transition.


\end{abstract}

\maketitle

\section{Introduction}

The standard cosmological model provides a remarkably successful description of a wide range of observations, yet some recent results suggest that it may be incomplete. In particular, local determinations of 
$H_0$ remain in tension with values inferred from early-time observations within the $\Lambda$CDM framework~\cite{Riess2022,Wong2020,Freedman2019,Pesce2020}. More recently, analyses including the baryon acoustic oscillation (BAO) measurements from DESI have suggested a preference for dynamical dark energy, further motivating extensions of the standard dark energy sector~\cite{Adame2025}. 
 
On the theoretical side, the situation is equally intriguing. Swampland arguments have been interpreted as a challenge to the construction of metastable de Sitter vacua in string theory \cite{Palti2019,Kallosh2019,Grana2021}. This has motivated renewed interest in scenarios where the dark-energy sector contains a negative cosmological constant or experiences a sign change over cosmic time. Low-redshift analyses with a negative cosmological constant were revisited in Ref.~\cite{Visinelli2019}. Related possibilities have also been discussed in connection with early JWST observations \cite{Adil2023,Menci2024}, with recent DESI BAO measurements \cite{Wang2024}, and with sign-switching dark-energy models \cite{Akarsu2020,Akarsu2021,Akarsu2023}. A complementary theoretical route, in which the sign of the cosmological constant changes as the Universe cools, has recently been proposed in Ref.~\cite{Nyergesy2025}. 

As is well known, no fundamental principle forbids a nonminimal coupling in the cosmological dark sector. We therefore explore this possibility and consider an interacting dark energy model in which dark matter and dark energy exchange energy through a rather general linear coupling (hereafter Linear Interacting Dark Energy, LIDE). In the absence of guidance from fundamental physics, a variety of phenomenological interaction terms 
$Q$ have been proposed. In some of these models, the region of parameter space preferred by current observations can yield an unphysical negative dark matter density $\rho_{\text{dm}}$ at late times, thereby violating the weak energy condition (WEC), $\rho_{\text{dm}} \geq 0$ \cite{Gavela2009,Marttens2020}. Here we argue that the full linear interaction \cite{first_linear,Wands}
\begin{equation} \label{eq:qgeneral}
Q = 3 H \left(\xi_1 \rho_{\Lambda} + \xi_2 \rho_{\rm dm}\right)
\end{equation}
provides a natural framework for describing energy exchange in the dark sector. In particular, it admits an interpretation analogous to a two-way reaction system, with the two dimensionless coefficients $\xi_1$ and $\xi_2$ governing opposite energy-transfer channels.  

Within this framework, our main goal is to examine whether it can accommodate a cosmological evolution in which the vacuum is effectively anti-de Sitter in the past and de Sitter at late times, while remaining compatible with current observations. To this end, we derive the analytic background solutions and perturbation equations, analyze viable cosmological evolutions, and perform a Markov Chain Monte Carlo (MCMC) analysis using \textcolor{black}{DES-Dovekie} supernovae, DESI BAO and CMB data. Our results yield $\xi_1 = 0.0083 \pm 0.0055$ and $\xi_2 =-0.0004 \pm 0.0004$ at 68\% confidence level, consistent with the physically motivated regime $\xi_1 > 0$ and $\xi_2 < 0$. In this regime, the model avoids the dark matter WEC violation while allowing the dark energy density $\rho_{\Lambda}$ to evolve from negative to positive values.

The paper is organized as follows. In Sec.~\ref{sec:model} we summarize the interacting model and its background solutions. Section~\ref{sec:evolution} discusses viable background evolutions, while linear perturbations are developed in Sec.~\ref{sec:perturbations}. Sec.~\ref{sec:data} presents the cosmological data sets used in our analysis and statistical methodology, and observational constraints are presented in Sec.~\ref{sec:constraints}. We conclude the paper by summarizing our main conclusions in Sec.~\ref{sec:conclusion}. Technical details of the analytic solutions are discussed in Appendix~\ref{app:background}.

\section{Cosmological model}
\label{sec:model}

\subsection{Linear interacting dark energy}

We consider a spatially flat FLRW spacetime with interacting dark matter and dark energy. Their balance equations are
\begin{align}
\nabla_{\mu} T^{\mu}_{\phantom{\mu}\nu\,({\rm dm})} &= Q_{\nu},
\\
\nabla_{\mu} T^{\mu}_{\phantom{\mu}\nu\,({\rm de})} &= -Q_{\nu},
\end{align}
where $Q_{\nu}$ accounts for the energy-momentum transfer between the dark components. At the background level this reduces to
\begin{align}
\dot{\rho}_{\rm dm} + 3 H \rho_{\rm dm} &= Q,
\label{eq:contdm}
\\
\dot{\rho}_{\Lambda} + 3 H \rho_{\Lambda} (1+w_{\Lambda}) &= -Q,
\label{eq:contde}
\end{align}
where $w_{\Lambda}$ is the dark-energy equation-of-state parameter.

We adopt the general linear interaction (\ref{eq:qgeneral}) 
with constant dimensionless couplings $\xi_1$ and $\xi_2$. The analytic background solutions can be written in closed form once one defines
\begin{equation}
\Delta = \xi_1^2 + \xi_2^2 + w_{\Lambda}^2 - 2\xi_1\xi_2 + 2\xi_1 w_{\Lambda} + 2\xi_2 w_{\Lambda}.
\label{eq:Delta}
\end{equation}
For $\Delta > 0$, the energy densities are given by
\begin{align}
\rho_{\rm dm} &= D_1 a^{3\lambda_1} + D_2 a^{3\lambda_2},
\label{eq:rhodmpos}
\\
\rho_{\Lambda} &= C_1 a^{3\lambda_1} + C_2 a^{3\lambda_2},
\label{eq:rholambdapos}
\end{align}
where
\begin{equation}
\lambda_{1,2} = \frac{1}{2}\left(-\xi_1 + \xi_2 - w_{\Lambda} - 2 \mp \sqrt{\Delta}\right),
\end{equation}
and the coefficients $C_i$ and $D_i$ are fixed by the present-day densities. Explicit expressions are listed in Appendix~\ref{app:background}.

The cases $\Delta = 0$ and $\Delta < 0$ can also be solved analytically, but they are much less attractive from a physical point of view. When $\Delta = 0$, logarithmic terms appear in the solutions and tend to drive the dark-matter density negative at some stage of the evolution. When $\Delta < 0$, oscillatory terms arise and, in general, again lead to unphysical negative matter density. For this reason, although we allow the sampler to explore the full parameter space, the physically interesting region is expected to lie predominantly in the range $\Delta > 0$.

\subsection{Physical interpretation of the couplings}

A useful way to think about Eq.~\eqref{eq:qgeneral} is by analogy with a two-component reaction system,
\begin{align}
\frac{d\rho_A}{dt} &= \alpha \rho_B - \beta \rho_A,
\\
\frac{d\rho_B}{dt} &= -\alpha \rho_B + \beta \rho_A.
\end{align}
Here $\alpha$ and $\beta$ represent transfer rates in opposite directions. When two components can exchange energy both ways, describing the system with two terms is more natural than forcing one coefficient to carry both directions through its sign. This motivates the interpretation of $\xi_1$ and $\xi_2$ as effective reaction rates. 
In the MCMC analysis we do not impose sign conditions a priori. Instead, we first explore the full parameter space and only afterwards examine which posterior regions satisfy the WEC for dark matter and which admit the anti-de Sitter to de Sitter transition.

\section{Background evolution}
\label{sec:evolution}

\begin{figure*}[t]
    \centering
    \includegraphics[width=0.88\textwidth]{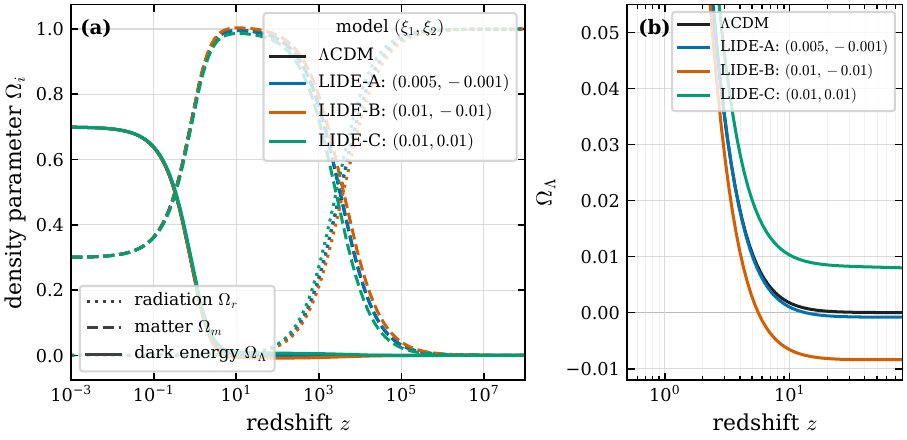}
    \caption{Evolution of the density parameters for $\Lambda$CDM and the three representative LIDE models defined in the text. Colors identify the model, while dotted, dashed and solid lines denote radiation, total matter and dark energy, respectively. The right panel enlarges the zero-crossing region. LIDE-A and LIDE-B undergo a negative-to-positive transition of $\rho_\Lambda$, whereas LIDE-C remains positive.}
    \label{fig:omega}
\end{figure*}

\begin{figure*}[t]
    \centering
    \includegraphics[width=0.88\textwidth]{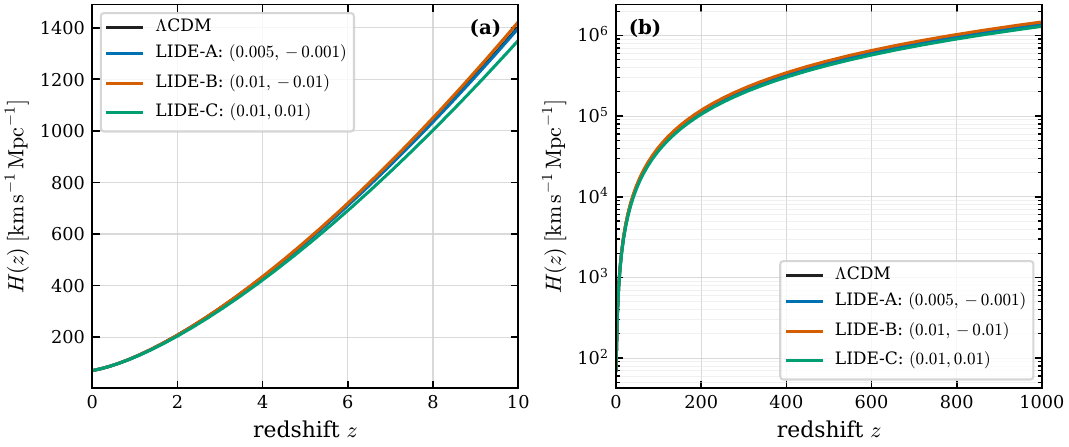}
    \caption{Hubble parameter for the same models as in Fig.~\ref{fig:omega}. The left panel covers $0\leq z\leq10$, while the right panel extends to $0\leq z\leq1000$ and uses a logarithmic vertical scale. The interacting models remain close to the $\Lambda$CDM background expansion.}
    \label{fig:hubble}
\end{figure*}

\begin{figure}[t]
    \centering
    \includegraphics[width=0.97\columnwidth]{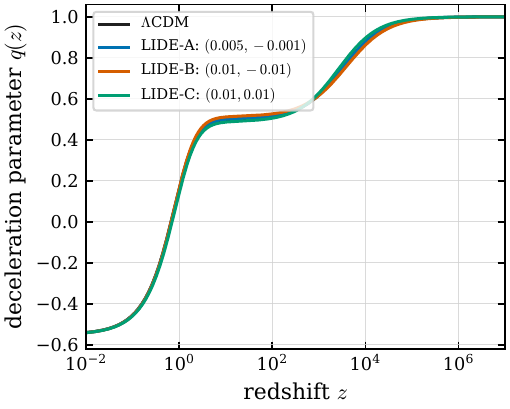}
    \caption{Deceleration parameter as a function of redshift. Despite the sign change in the dark-energy density of LIDE-A and LIDE-B, the radiation-, matter- and acceleration-dominated stages remain close to those of $\Lambda$CDM.}
    \label{fig:decel}
\end{figure}

Before turning to the data, it is useful to look at the background cosmological evolution implied by the model. Throughout Figs.~\ref{fig:omega}--\ref{fig:decel}, we fix $H_0=70\,\mathrm{km\,s^{-1}\,Mpc^{-1}}$, $\Omega_{m0}=0.3$, $\Omega_{b0}=0.0493$, $\Omega_{r0}=9.27\times10^{-5}$ and $w_\Lambda=-1$. We compare $\Lambda$CDM with three illustrative LIDE models: LIDE-A $(\xi_1,\xi_2)=(0.005,-0.001)$, LIDE-B $(0.01,-0.01)$, and LIDE-C $(0.01,0.01)$. The first two examples preserve a positive dark-matter density and exhibit a negative-to-positive transition of $\rho_\Lambda$ at $z\simeq13.3$ and $z\simeq5.6$, respectively; LIDE-C is included as a non-transition comparison.

Figure~\ref{fig:omega} shows the corresponding evolution of the density parameters. The left panel displays the full evolution, while the right panel isolates the regime where the dark energy density crosses zero. The sign change is therefore generated dynamically by the interaction rather than imposed as a redshift-dependent parametrization.


Figures~\ref{fig:hubble} and \ref{fig:decel} show the expansion history. Even in the transition scenario, the Hubble rate and the deceleration parameter remain close to their $\Lambda$CDM counterparts at the background level. The model still reproduces the familiar radiation-dominated, matter-dominated and late-time accelerated eras. The main difference is therefore not a dramatic reshaping of the expansion history, but the possibility of a sign change in the dark-energy density.

These figures already suggest an important point: the signs of $\xi_1$ and $\xi_2$ do not strongly affect the homogeneous expansion history by themselves. In practice, the parameter $w_{\Lambda}$ plays a more visible role in shaping the background evolution. This is why observational constraints discussed in secs.~\ref{sec:data} and \ref{sec:constraints} are essential for identifying the preferred region of the interaction parameter space.

\section{Linear perturbations of the interacting dark sector}
\label{sec:perturbations}

In this section we summarize the covariant perturbative formulation of the interacting dark-energy model. The energy--momentum balance equations for the interacting components are
\begin{align}
\nabla_{\mu}T^{\mu\nu}_{\text{dm}} &= Q^{\nu},
\label{eq:pert_balance_m}
\\
\nabla_{\mu}T^{\mu\nu}_{x} &= -Q^{\nu},
\label{eq:pert_balance_x}
\end{align}
where
\begin{equation}
T^{\mu\nu}_{A}=\rho_A u_A^{\mu}u_A^{\nu}+p_A\left(g^{\mu\nu}+u_A^{\mu}u_A^{\nu}\right),
\label{eq:pert_energy_momentum_tensor}
\end{equation}
with $x$ indicating dark energy in this section. Baryons are conserved separately. In general, each component has its own four-velocity. The interaction vector can be decomposed into parts parallel and orthogonal to the dark-matter four-velocity $u^{\mu}$,
\begin{equation}
Q^{\mu}=Q u^{\mu}+\bar{Q}^{\mu},
\label{eq:pert_q_decomposition}
\end{equation}
with
\begin{align}
-u_{\mu}Q^{\mu} &= Q, \\
u_{\mu}\bar{Q}^{\mu} &=0, \\
u_{\mu}u^{\mu} &=-1.
\end{align}
In the comoving frame, $u^{\mu}=a^{-1}(1,\boldsymbol{0})$. Using the FLRW metric in Eqs.~\eqref{eq:pert_balance_m} and \eqref{eq:pert_balance_x}, the background continuity equations are
\begin{align}
\rho_\text{dm}' + 3\mathcal{H}\rho_\text{dm} &= aQ,
\label{eq:pert_background_m}
\\
\rho_x' + 3\mathcal{H}\rho_x(1+w_x) &= -aQ,
\label{eq:pert_background_x}
\end{align}
where $\mathcal{H}=aH$, the prime denotes derivative with respect to conformal time, $w_x$ is the dark-energy equation-of-state parameter, and $\rho_\text{dm}$ and $\rho_x$ are the dark-matter and dark-energy densities. 

We perturb the background metric in the longitudinal gauge,
\begin{equation}
ds^2=a^2(\eta)\left[-(1+2\phi)d\eta^2+(1-2\psi)\delta_{ij}dx^i dx^j\right],
\label{eq:pert_longitudinal_metric}
\end{equation}
where $\phi(\boldsymbol{x},\eta)$ can be identified with the Newtonian gravitational potential in the appropriate limit. 
Perturbing the interaction term in Eq.~\eqref{eq:pert_q_decomposition}, one obtains
\begin{align}
\delta Q_0 &= -a(\delta Q+Q\phi),
\label{eq:pert_delta_q_0}
\\
\delta Q_i &= a\partial_i(Qv + \delta q),
\label{eq:pert_delta_q_i}
\end{align}
where $\delta\bar{Q}_i=\partial_i\delta q$ and $v$ is the dark-matter velocity potential. For each component $A$, the perturbed balance equations are
\begin{widetext}
\begin{align}
\delta\rho_A' + 3\mathcal{H}(\delta\rho_A+\delta p_A)
-3(\rho_A+p_A)\psi' - k^2(\rho_A+p_A)v_A
=a(\delta Q_A+Q_A\phi),
\label{eq:pert_balance_density_general}
\\
\left[(\rho_A+p_A)v_A\right]'
+4\mathcal{H}(\rho_A+p_A)v_A
+(\rho_A+p_A)\phi+\delta p_A
-\frac{2}{3}k^2\pi_A
=a(Q_Av+\delta q_A),
\label{eq:pert_balance_velocity_general}
\end{align}
\end{widetext}
where we have moved to Fourier space, so that $\nabla^2=-k^2$.

\subsection{Dark matter perturbations}

For pressureless dark matter, Eqs.~\eqref{eq:pert_balance_density_general}-\eqref{eq:pert_balance_velocity_general} give
\begin{equation}
\delta_\text{dm}' - 3\psi' + V_\text{dm}
= \frac{aQ}{\rho_\text{dm}}(\phi-\delta_\text{dm})+\frac{a\delta Q}{\rho_\text{dm}},
\label{eq:pert_matter_delta}
\end{equation}
\begin{equation}
V_\text{dm}' + \mathcal{H}V_\text{dm} - k^2\phi
=-\frac{ak^2}{\rho_\text{dm}}\delta q,
\label{eq:pert_matter_velocity}
\end{equation}
where $V_\text{dm}=-k^2 v$ and $\delta_{A}=\delta\rho_{A}/\rho_{A}$. The background equation \eqref{eq:pert_background_m} has been used to eliminate $\rho_\text{dm}'$.

\subsection{Dark energy perturbations}

For dark energy, 
the perturbation equations become
\begin{widetext}
\begin{align}
\delta_x'-(1+w_x)(3\psi'-V_x)
=-\frac{aQ}{\rho_x}(\phi-\delta_x)-\frac{a\delta Q}{\rho_x},
\label{eq:pert_de_delta}
\\
(1+w_x)\left[V_x'+\mathcal{H}(1-3w_x)V_x
-k^2\phi\right]
-k^2w_x\delta_x
=-\frac{aQ}{\rho_x}\left[V_{dm}-(1+w_x)V_x\right]
+\frac{ak^2}{\rho_x}\delta q,
\label{eq:pert_de_velocity}
\end{align}
\end{widetext}
where Eq.~\eqref{eq:pert_background_x} has been used to eliminate $\rho_x'$.

The generalized Poisson equation is given by
\begin{equation}
-k^2\psi=\frac{a^2}{2}\left[\rho_m\delta_m+\rho_x\delta_x+\frac{3\mathcal{H}}{k^2}\left(\rho_m+\rho_x(1+w_x)\right)V\right],
\label{eq:pert_general_poisson}
\end{equation}
where $\rho_m = \rho_\text{dm} + \rho_b$, and the total fluid velocity is
\begin{equation}
V=\frac{\rho_m}{\rho_m+\rho_x(1+w_x)}V_m
+\frac{(1+w_x)\rho_x}{\rho_m+\rho_x(1+w_x)}V_x,
\label{eq:pert_total_velocity}
\end{equation}
with $\rho_m V_m = \rho_\text{dm} V_\text{dm} + \rho_b V_b$.
In what follows, we assume vanishing intrinsic momentum transfer,
\begin{equation}
\delta q=0,
\label{eq:pert_delta_q_zero}
\end{equation}
and set $\phi=\psi$, corresponding to null anisotropic stress. 


For $w_x=-1$, $V = V_m$, $V_x$ decouples and the system (\ref{eq:pert_matter_delta})-(\ref{eq:pert_general_poisson}), combined with the balance equations for baryons, is closed: $\delta_\text{dm}$, $V_\text{dm}$ and $\phi$ are determined once the background cosmological model is specified. In the sub-horizon limit $k\gg aH$, the system reduces to
\begin{align}
\delta_\text{dm}' + V_\text{dm} &\simeq -\frac{aQ}{\rho_\text{dm}}\delta_\text{dm},
\label{eq:pert_wminusone_delta_m}
\\
V_\text{dm}' + \mathcal{H}V_\text{dm} - k^2\phi &\simeq 0,
\label{eq:pert_wminusone_velocity_m}
\\
-k^2\phi &\simeq \frac{a^2}{2}\rho_m\delta_m.
\label{eq:pert_wminusone_poisson}
\end{align}
In this limit, the dark energy and energy-transfer perturbations $\delta \rho_x$ and $\delta Q$ are negligible. 

\label{eq:pert_covariant_q}

\section{Data Analysis}
\label{sec:data}

\subsection{DES-Dovekie Type Ia supernovae}

Type Ia supernovae provide precise measurements of relative cosmological distances through the distance modulus
\begin{equation}
\mu = m_B - M_B\,,
\end{equation}
where $m_B$ is the apparent magnitude and $M_B$ is the absolute magnitude. The theoretical prediction is
\begin{equation}
\mu(z) = 5 \log_{10}\left(\frac{D_L(z)}{1\,\mathrm{Mpc}}\right) + 25,
\end{equation}
where $D_L(z)$ is the luminosity distance. We use the DES-Dovekie compilation, an updated reanalysis of the five-year Dark Energy Survey supernova sample (DES-SN5YR) incorporating improved photometric cross-calibration and light-curve modeling. The final cosmological sample contains 1820 supernovae, comprising 1623 likely Type Ia supernovae from DES and 197 spectroscopically confirmed low-redshift Type Ia supernovae from complementary surveys \cite{DES:2025sig}. In the likelihood analysis, the absolute-magnitude offset is analytically marginalized over, such that the supernova data constrain only relative luminosity distances.

\subsection{DESI BAO data}

BAO measurements act as a standard ruler and constrain combinations of the Hubble parameter, the transverse comoving distance and the sound horizon at the drag epoch. We use the DESI DR2 BAO measurements \cite{Karim2025a,Karim2025b}, with the likelihood implemented in \texttt{Cobaya}. The theoretical predictions are obtained from the background solutions of the interacting model.

\subsection{Cosmic microwave background}

We include cosmic microwave background (CMB) measurements from the Planck Public Release 4 (PR4), obtained with the NPIPE processing pipeline \cite{Planck:2020olo}. The NPIPE reprocessing combines data from the Planck Low and High Frequency Instruments and incorporates improved calibration and treatment of instrumental systematics, resulting in reduced noise and improved consistency between frequency channels.

For the CMB likelihood, we use the Planck PR4 HiLLiPoP and LoLLiPoP likelihoods \cite{Tristram:2023haj}. At high multipoles, we adopt the HiLLiPoP $TTTEEE$ likelihood, which uses the temperature and polarization auto- and cross-spectra constructed from the 100, 143 and 217 GHz Planck frequency maps over the range $30\lesssim\ell\lesssim2500$. At low multipoles, the large-scale $E$-mode polarization information is included through the LoLLiPoP likelihood, while the low-$\ell$ temperature contribution is described by the Commander likelihood. This combination allows the CMB data to constrain the physical dark-matter and baryon densities, the primordial power spectrum and the acoustic scale, thereby providing complementary information to the late-time distance probes considered in our analysis.

\begin{table}[t]
\caption{Reference values and prior ranges adopted in the MCMC analysis. All priors are uniform.}
\label{tab:priors}
\centering
\small
\begin{tabular}{lccc}\hline\hline
Parameter & Reference value & Min & Max \\
\hline
$H_0\, (\mathrm{km\,s^{-1}\,Mpc^{-1}})$ & 70.0 & 40.0 & 100.0 \\
$\Omega_{m0}$ & 0.3 & 0.0 & 1.0 \\
$\Omega_b h^2$ & 0.0224 & 0.020 & 0.024 \\
$\ln(10^{10}A_s)$ & 3.05 & 1.61 & 3.91 \\
$n_s$ & 0.965 & 0.8 & 1.2 \\
$\tau_{\rm reio}$ & 0.055 & 0.01 & 0.8 \\
$\xi_1$ & 0.0 & -1.0 & 1.0 \\
$\xi_2$ & 0.0 & -1.0 & 1.0 \\
\hline\hline
\end{tabular}
\end{table}

\begin{figure*}[t]
    \centering
    \includegraphics[width=0.86\textwidth]{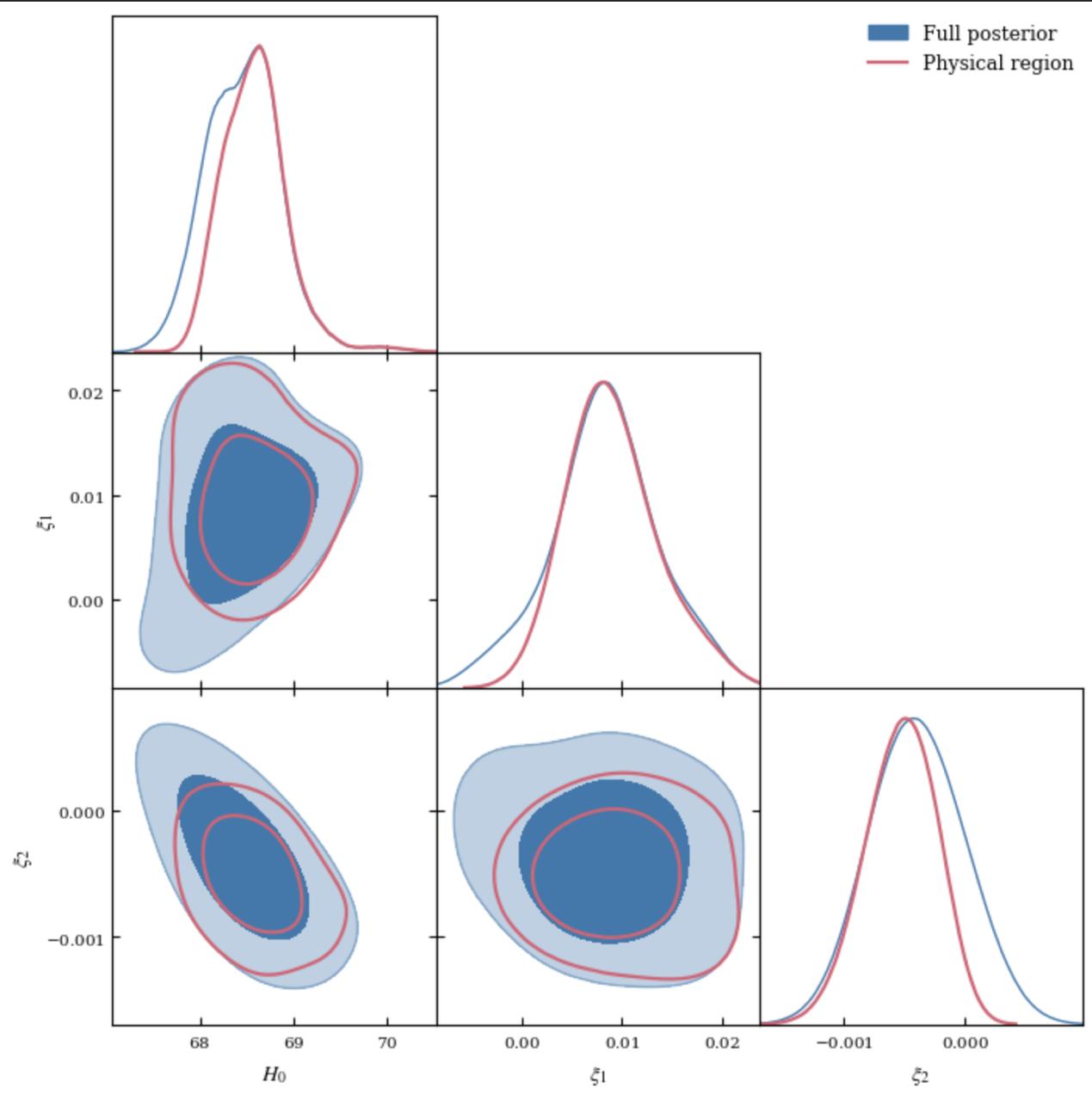}
    \caption{Posterior distribution for the LIDE model using DES-Dovekie supernovae, DESI BAO and Planck CMB data. Blue shows the full chain, and magenta the physical region preserving the weak energy condition for dark matter and admitting a transition from an anti-de Sitter-like phase to a de Sitter-like phase.} 
    \label{fig:posterior}
\end{figure*}

\begin{table*}[t]
\caption{Marginalized constraints for the full LIDE model and for the region preserving the weak energy condition for dark matter and allowing the anti-de Sitter to de Sitter transition. Uncertainties correspond to 68\% confidence level.}
\label{tab:constraints}
\centering
\begin{tabular}{l|cc|cc}
\hline\hline
\multicolumn{1}{c|}{\multirow{2}{*}{$\qquad$Parameter$\qquad$}} & \multicolumn{2}{c|}{$\qquad$LIDE (full posterior)$\qquad$}   & \multicolumn{2}{c}{$\qquad$LIDE (physical region)$\qquad$}  \\ \cline{2-5} 
\multicolumn{1}{c|}{}                           & best fit    & mean$\pm$std           & best fit    & mean$\pm$std          \\ \hline
$\omega_{\rm b}$                                & $0.02218$   & $0.02240\pm0.00020$    & $0.02218$   & $0.02237\pm0.00019$   \\
$\Omega_{\rm m}$                                & $0.307$     & $0.306\pm0.005$        & $0.307$     & $0.305\pm0.005$       \\
$H_0$                                           & $68.19$     & $68.46\pm0.41$         & $68.19$     & $68.58\pm0.35$        \\
$\tau_{\rm reio}$                               & $0.0478$    & $0.0517\pm0.0109$      & $0.0478$    & $0.0501\pm0.0097$     \\
$n_{\rm s}$                                     & $0.9664$    & $0.9658\pm0.0038$      & $0.9664$    & $0.9658\pm0.0039$     \\
$\xi_1$                                         & $0.0053$    & $0.0083\pm0.0055$      & $0.00532$   & $0.0090\pm0.0045$     \\
$\xi_2$                                         & $-0.0005$   & $-0.0004\pm0.0004$     & $-0.0005$   & $-0.0005\pm0.0003$    \\ \hline\hline
\end{tabular}
\end{table*}

\subsection{Calibration and nuisance parameters}

The DES-Dovekie supernova likelihood is analytically marginalized over the absolute-magnitude offset $M_B$. Since $M_B$ is fully degenerate with the overall normalization of the luminosity distance, and hence with $H_0$, the supernova likelihood does not provide an independent constraint on the absolute distance scale. Consequently, $M_B$ is not included as a sampled nuisance parameter in our analysis. The Hubble constant $H_0$ is nevertheless varied in the joint analysis and is constrained by the complementary cosmological probes.

The absolute distance scale in our joint analysis is instead determined through the BAO measurements. Their theoretical prediction depends on the sound horizon at the drag epoch,
\begin{equation}
r_d = \int_{z_d}^{\infty} \frac{c_s(z)}{H(z)}\, dz,
\end{equation}
where $c_s(z)$ is the sound speed in the baryon--photon plasma. We therefore include $\Omega_b h^2$ as an additional parameter entering the computation of $r_d$.

The nuisance parameters associated with the Planck PR4 HiLLiPoP and LoLLiPoP likelihoods, including calibration and foreground contributions, are varied using the default likelihood prescriptions and priors implemented in \texttt{Cobaya}.

The MCMC analysis is performed with \texttt{Cobaya} \cite{Torrado2021}, using its default sampler \cite{LewisBridle2002,Lewis2013}. We consider in our runs the usual $\Lambda$CDM parameters and the interacting parameters: $H_0$, $\Omega_{m0}$, $\Omega_b h^2$, $\ln(10^{10}A_s)$, $n_s$, $\tau_{\rm reio}$, $\xi_1$, and $\xi_2$. The dark energy equation-of-state parameter is fixed as $w_{\Lambda}=-1$, which guarantees that this component is strictly homogeneous at observational scales, as shown above. The priors are uniform and listed in Table~\ref{tab:priors}. We monitor convergence with the Gelman--Rubin statistic and require $R-1 < 0.01$ \cite{GelmanRubin1992}.

\section{Observational constraints}
\label{sec:constraints}

After sampling the full parameter space, we inspect two nested cases: (i) the full LIDE posterior, with no physical cut beyond the model definition; (ii) the posterior region satisfying the WEC for dark matter and that also allows a transition from negative to positive dark energy.
For $w_{\Lambda} = -1$, the latter is defined by the condition
\begin{equation}
\xi_2 < - \frac{\Omega_{\Lambda0}}{\Omega_{\text{dm}0}} \xi_1 + \frac{\Omega_{\Lambda0}}{\Omega_{\Lambda0} + \Omega_{\text{dm0}}},
\end{equation}
together with $\xi_1 \geq 0$ and $\xi_2 < 0$, where $\Omega_{\text{dm}0} = \Omega_{m0} - \Omega_{b0}$.
Figure~\ref{fig:posterior} shows the posterior distributions. 

We have also constrained the WEC-preserving region and the transition region separately, and find that they overlap almost completely. In other words, within the parameter space preferred by the data, once the clearly unphysical region associated with the dark-matter sector is excluded, the transition scenario emerges almost automatically. This result should nevertheless be interpreted with some caution, as the  overlap may partly reflect the structure of the model and the cuts imposed in the analysis. Even so, the near coincidence of the two regions suggests that the transition may be a generic feature of the observationally viable parameter space rather than a finely tuned possibility.


The marginalized constraints are summarized in Table~\ref{tab:constraints}. The best-fit point lies fairly close to the standard picture, with $\Omega_{m0}\approx 0.3$.
More importantly for our purposes, the data show a clear tendency toward $\xi_1 > 0$ and $\xi_2 < 0$. 
The constraints remain broad, which is not surprising given the well-known degeneracies introduced by interacting dark-energy models. Still, the preferred region is physically meaningful and supports the transition scenario.


The present data do not decisively favor this interacting model over $\Lambda$CDM. At the same time, they do not rule it out, and they are perfectly consistent with the physically motivated region in which an anti-de Sitter-like past evolves into a de Sitter-like present. This places the model in interesting contact with recent discussions of negative cosmological constants and sign-switching dark energy in the literature \cite{Visinelli2019,Adil2023,Menci2024,Wang2024,Akarsu2020,Akarsu2021,Akarsu2023,Nyergesy2025}. Our scenario differs from those approaches because the sign change is not imposed in the vacuum sector, but emerges dynamically from the interaction with dark matter.

\section{Conclusions}
\label{sec:conclusion}

We have studied the general linear interacting dark-energy model at the background and perturbative levels and shown that it can accommodate a transition from a negative to a positive dark-energy density. In that sense, the model offers a simple dynamical realization of an anti-de Sitter to de Sitter transition. The mechanism is especially natural in the region $\xi_1 > 0$ and $\xi_2 < 0$, where 
the dark-matter component is always positive, in agreement with the WEC.

Using DES-Dovekie supernovae, DESI BAO and Planck PR4/NPIPE CMB data, we found that the preferred posterior region shows good agreement with this physically motivated picture. In fact, after excluding the  unphysical dark-matter sector, the remaining posterior almost entirely supports the transition scenario. The interaction parameters in that region are constrained to
$\xi_1 = 0.0090 \pm 0.0045$ and $\xi_2 = -0.0005 \pm 0.0003$
at 68\% confidence level.

The picture that emerges is suggestive. Recent works have shown that negative cosmological constants and sign-changing dark-energy sectors remain phenomenologically relevant in light of low-redshift observations, JWST-inspired discussions, DESI results and string-theory-motivated considerations \cite{Palti2019,Visinelli2019,Adil2023,Menci2024,Wang2024,Nyergesy2025}. Our analysis adds to that discussion by showing that a dark-sector interaction can provide a concrete and observationally allowed route from an anti-de Sitter-like phase to the de Sitter-like Universe observed today.

\appendix

\section{Background solutions of the LIDE model}
\label{app:background}

For completeness, we summarize the analytic background solutions. Starting from Eqs.~\eqref{eq:contdm} and \eqref{eq:contde}, and using $H = \dot a/a$, we obtain
\begin{align}
 a\frac{d\rho_{\rm dm}}{da} &= 3(\xi_2-1)\rho_{\rm dm} + 3\xi_1\rho_{\Lambda},
 \\
 a\frac{d\rho_{\Lambda}}{da} &= -3\xi_2\rho_{\rm dm} - 3(\xi_1 + 1 + w_{\Lambda})\rho_{\Lambda}.
\end{align}
With the change of variable $t = \ln a$, the system can be written as
\begin{equation}
\frac{d}{dt}
\begin{pmatrix}
\rho_{\rm dm} \\
\rho_{\Lambda}
\end{pmatrix}
= 3
\begin{pmatrix}
\xi_2-1 & \xi_1 \\
-\xi_2 & -(1+w_{\Lambda}+\xi_1)
\end{pmatrix}
\begin{pmatrix}
\rho_{\rm dm} \\
\rho_{\Lambda}
\end{pmatrix}.
\end{equation}
The nature of the solution depends on the discriminant $\Delta$ defined in Eq.~\eqref{eq:Delta}.

\subsection{$\Delta > 0$}

This is the physically most relevant case. When $\Delta > 0$, the eigenvalues are real and distinct,
\begin{equation}
\lambda_{1,2} = \frac{1}{2}\left(-\xi_1 + \xi_2 - w_{\Lambda} - 2 \mp \sqrt{\Delta}\right),
\end{equation}
and the solutions can be written as in Eqs.~\eqref{eq:rhodmpos} and \eqref{eq:rholambdapos}, with
\begin{align}
D_1 &= -\frac{\xi_1 + \xi_2 + w_{\Lambda} - \sqrt{\Delta}}{2\sqrt{\Delta}}\rho_{{\rm dm}0}
      -\frac{\xi_1}{\sqrt{\Delta}}\rho_{\Lambda 0},
\\
D_2 &= \frac{\xi_1 + \xi_2 + w_{\Lambda} + \sqrt{\Delta}}{2\sqrt{\Delta}}\rho_{{\rm dm}0}
      +\frac{\xi_1}{\sqrt{\Delta}}\rho_{\Lambda 0},
\\
C_1 &= \frac{\xi_2}{\sqrt{\Delta}}\rho_{{\rm dm}0}
      +\frac{\xi_1 + \xi_2 + w_{\Lambda} + \sqrt{\Delta}}{2\sqrt{\Delta}}\rho_{\Lambda 0},
\\
C_2 &= -\frac{\xi_2}{\sqrt{\Delta}}\rho_{{\rm dm}0}
      -\frac{\xi_1 + \xi_2 + w_{\Lambda} - \sqrt{\Delta}}{2\sqrt{\Delta}}\rho_{\Lambda 0}.
\end{align}

\subsection{$\Delta = 0$}

When $\Delta=0$, the matrix has a degenerate eigenvalue and the Jordan form introduces logarithmic corrections,
\begin{align}
\rho_{\rm dm} &= D_1 a^{3\lambda} + D_2 a^{3\lambda}\ln a,
\\
\rho_{\Lambda} &= C_1 a^{3\lambda} + C_2 a^{3\lambda}\ln a,
\end{align}
with
\begin{equation}
\lambda = \frac{1}{2}\left(-\xi_1 + \xi_2 - w_{\Lambda} -2\right).
\end{equation}
Unless the logarithmic coefficients vanish in a fine-tuned way, these solutions typically become problematic for dark matter at some stage of the evolution.

\subsection{$\Delta < 0$}

For $\Delta < 0$, the eigenvalues are complex and the solutions become oscillatory. Defining
\begin{equation}
\lambda_* = \frac{1}{2}\left(-\xi_1 + \xi_2 - w_{\Lambda} -2\right),
\end{equation}
we obtain
\begin{align}
\rho_{\rm dm} &=  a^{3\lambda_*}
\left[ D_1\cos\!\left(\frac{3}{2}\sqrt{-\Delta}\ln a\right)
+ D_2\sin\!\left(\frac{3}{2}\sqrt{-\Delta}\ln a\right)\right],
\\
\rho_{\Lambda} &=  a^{3\lambda_*}
\left[ C_1\cos\!\left(\frac{3}{2}\sqrt{-\Delta}\ln a\right)
+ C_2\sin\!\left(\frac{3}{2}\sqrt{-\Delta}\ln a\right)\right],
\end{align}
where
\begin{align}
C_1 &= \rho_{\Lambda0},
\\
C_2 &= \frac{2\xi_2}{\sqrt{-\Delta}}(v_*\rho_{\Lambda0} - \rho_{dm0}),
\\
D_1 &= \rho_{dm0},
\\
D_2 &= \frac{\sqrt{-\Delta}}{2\xi_2}\rho_{\Lambda0}+v_*C_2,
\end{align}
with \begin{equation}
v_* = -\frac{\xi_1 + \xi_2 + w_{\Lambda}}{2\xi_2}.
\end{equation}
Although mathematically allowed, this regime is not attractive phenomenologically because the oscillations tend to generate negative values for the dark matter density.

\section*{Acknowledgments}

YZ is supported by a PhD grant from the Coordena\c{c}\~ao de Aperfei\c{c}oamento de Pessoal de N\'ivel Superior (CAPES). JSA is supported by Conselho Nacional de Desenvolvimento Científico e Tecnológico (CNPq) grants Nos. 307683/2022-2 and 448158/2025-6 and Funda\c{c}\~ao de Amparo \`a Pesquisa do Estado do Rio de Janeiro (FAPERJ) grant No. 299312 (2023). SC is supported by CNPq with grant 308518/2023-3. RvM is supported by CNPq with grant 311114/2026-1. The development of this work was aided by the National Observatory Data Center (CPDON).


\begin{thebibliography}{99}

\bibitem{Riess2022}
A.~G.~Riess \textit{et al.},
``A comprehensive measurement of the local value of the Hubble constant with 1 km s$^{-1}$ Mpc$^{-1}$ uncertainty from the Hubble Space Telescope and the SH0ES team,''
Astrophys. J. Lett. \textbf{934}, L7 (2022).

\bibitem{Wong2020}
K.~C.~Wong \textit{et al.},
``H0LiCOW XIII: A 2.4 per cent measurement of $H_0$ from lensed quasars: 5.3$\sigma$ tension between early- and late-Universe probes,''
Mon. Not. Roy. Astron. Soc. \textbf{498}, 1420 (2020).

\bibitem{Freedman2019}
W.~L.~Freedman \textit{et al.},
``The Carnegie-Chicago Hubble Program VIII: An independent determination of the Hubble constant based on the tip of the red giant branch,''
Astrophys. J. \textbf{882}, 34 (2019).

\bibitem{Pesce2020}
D.~W.~Pesce \textit{et al.},
``The Megamaser Cosmology Project XIII: Combined Hubble constant constraints,''
Astrophys. J. Lett. \textbf{891}, L1 (2020).

\bibitem{Adame2025}
A.~G.~Adame \textit{et al.},
``DESI 2024 VI: Cosmological constraints from the measurements of baryon acoustic oscillations,''
JCAP \textbf{02}, 021 (2025).

\bibitem{Palti2019}
E.~Palti,
``The swampland: introduction and review,''
Fortschr. Phys. \textbf{67}, 1900037 (2019).

\bibitem{Kallosh2019}
R.~Kallosh, A.~Linde, E.~McDonough, and M.~Scalisi,
``dS vacua and the swampland,''
JHEP \textbf{03}, 134 (2019).

\bibitem{Grana2021}
M.~Gra\~na and A.~Herr\'aez,
``The swampland conjectures: a bridge from quantum gravity to particle physics,''
Universe \textbf{7}, 273 (2021).

\bibitem{Visinelli2019}
L.~Visinelli, S.~Vagnozzi, and U.~Danielsson,
``Revisiting a negative cosmological constant from low-redshift data,''
Symmetry \textbf{11}, 1035 (2019).

\bibitem{Adil2023}
S. A. Adil, U. Mukhopadhyay, A. A. Sen, and S. Vagnozzi, “Dark energy in light of the early JWST observations: case for a negative cosmological constant?”, JCAP {\bf 10}, 072 (2023).

\bibitem{Menci2024}
N. Menci, S. A. Adil, U. Mukhopadhyay, A. A. Sen, and S. Vagnozzi, “Negative cosmological constant in the dark energy sector: tests from JWST photometric and spectroscopic observations of high-redshift galaxies”, JCAP {\bf 07}, 072 (2024).

\bibitem{Wang2024}
H. Wang, Z.-Y. Peng, and Y.-S. Piao, “Can recent DESI BAO measurements accommodate a negative cosmological constant?”, Phys. Rev. D {\bf 111}, L061306 (2025).

\bibitem{Akarsu2020}
O.~Akarsu, J.~D.~Barrow, L.~A.~Escamilla, and J.~A.~Vazquez,
``Graduated dark energy: observational hints of a spontaneous sign switch in the cosmological constant,''
Phys. Rev. D \textbf{101}, 063528 (2020).

\bibitem{Akarsu2021}
O.~Akarsu, S.~Kumar, E.~Oz\"ulker, and J.~A.~Vazquez,
``Relaxing cosmological tensions with a sign switching cosmological constant,''
Phys. Rev. D \textbf{104}, 123512 (2021).

\bibitem{Akarsu2023}
O.~Akarsu, S.~Kumar, E.~Oz\"ulker, J.~A.~Vazquez, and A.~Yadav,
``Relaxing cosmological tensions with a sign switching cosmological constant: improved results with Planck, BAO, and Pantheon data,''
Phys. Rev. D \textbf{108}, 023513 (2023).

\bibitem{Nyergesy2025}
E. N. Nyergesy, I. G. Márián, A. Trombettoni, and I. Nándori, “From negative to positive cosmological constant through decreasing temperature of the Universe: connection with string theory and spacetime foliation results”, Phys. Lett. B {\bf 876}, 140380 (2026).

\bibitem{Gavela2009}
M.~B.~Gavela, D.~Hernandez, L.~Lopez Honorez, O.~Mena, and S.~Rigolin,
``Dark coupling'',
JCAP \textbf{07}, 034 (2009); Erratum ibid. {\bf 05}, E01 (2010).

\bibitem{Marttens2020}
R.~von Marttens, H.~A.~Borges, S.~Carneiro, J.~S.~Alcaniz, and W.~Zimdahl,
``Unphysical properties in a class of interacting dark energy models,''
Eur. Phys. J. C \textbf{80}, 1110 (2020).

\bibitem{first_linear} H. Mohseni Sadjadi and M. Alimohammadi, ``Cosmological coincidence problem in interacting dark energy models'', Phys. Rev. D {\bf 74}, 103007 (2006).


\bibitem{Wands} Chakkrit Kaeonikhom {\it et al.}, ``Observational constraints on interacting vacuum energy with linear interactions", JCAP {\bf 01}, 042 (2023).



\bibitem{Karim2025a}
M. Abdul Karim {\it et al.}, “DESI DR2 results. I. Baryon acoustic oscillations from the Lyman alpha forest”, Phys. Rev. D {\bf 112}, 083514 (2025).

\bibitem{Karim2025b}
M. Abdul Karim {\it et al.}, “DESI DR2 results. II. Measurements of baryon acoustic oscillations and cosmological constraints”, Phys. Rev. D {\bf 112}, 083515 (2025).



\bibitem{Torrado2021}
J.~Torrado and A.~Lewis,
``Cobaya: code for Bayesian analysis of hierarchical physical models,''
JCAP \textbf{05}, 057 (2021).

\bibitem{LewisBridle2002}
A.~Lewis and S.~Bridle,
``Cosmological parameters from CMB and other data: a Monte Carlo approach,''
Phys. Rev. D \textbf{66}, 103511 (2002).

\bibitem{Lewis2013}
A.~Lewis,
``Efficient sampling of fast and slow cosmological parameters,''
Phys. Rev. D \textbf{87}, 103529 (2013).

\bibitem{GelmanRubin1992}
A.~Gelman and D.~B.~Rubin,
``Inference from iterative simulation using multiple sequences,''
Statist. Sci. \textbf{7}, 457 (1992).


\bibitem{DES:2025sig}
B.~Popovic \textit{et al.} [DES],
``The Dark Energy Survey supernova program: a reanalysis of cosmology results and evidence for evolving dark energy with an updated Type Ia supernova calibration,''
Mon. Not. Roy. Astron. Soc. \textbf{548}, stag632 (2026).

\bibitem{Planck:2020olo}
Y.~Akrami \textit{et al.} [Planck],
``$Planck$ intermediate results. LVII. Joint Planck LFI and HFI data processing,''
Astron. Astrophys. \textbf{643}, A42 (2020).

\bibitem{Tristram:2023haj}
M.~Tristram \textit{et al.},  
``Cosmological parameters derived from the final Planck data release (PR4),''
Astron. Astrophys. \textbf{682}, A37 (2024).

\end{thebibliography}
\end{document}